\documentclass[superscriptaddress]{revtex4-2}
\pdfoutput=1
\usepackage[utf8]{inputenc}
\usepackage{graphicx}
\usepackage{xcolor}

\begin{document}

\title{A chiral topological add-drop filter for integrated quantum photonic circuits}

\author{M. Jalali Mehrabad}
\email[]{mjalalimehrabad1@sheffield.ac.uk}
\affiliation{Department of Physics and Astronomy, University of Sheffield, Sheffield S3 7RH, UK}

\author{A.P. Foster}
\email[]{andrew.foster@sheffield.ac.uk}
\affiliation{Department of Physics and Astronomy, University of Sheffield, Sheffield S3 7RH, UK}

\author{N.J. Martin}
\affiliation{Department of Physics and Astronomy, University of Sheffield, Sheffield S3 7RH, UK}

\author{R. Dost}
\affiliation{Department of Physics and Astronomy, University of Sheffield, Sheffield S3 7RH, UK}

\author{E. Clarke}
\affiliation{EPSRC National Epitaxy Facility, University of Sheffield, Sheffield S1 4DE, UK}

\author{P.K. Patil}
\affiliation{EPSRC National Epitaxy Facility, University of Sheffield, Sheffield S1 4DE, UK}

\author{M.S. Skolnick}
\affiliation{Department of Physics and Astronomy, University of Sheffield, Sheffield S3 7RH, UK}

\author{L.R. Wilson}
\affiliation{Department of Physics and Astronomy, University of Sheffield, Sheffield S3 7RH, UK}

\begin{abstract}
The integration of quantum emitters within topological nano-photonic devices opens up new avenues for the control of light-matter interactions at the single photon level. Here, we realise a spin-dependent, chiral light-matter interface using individual semiconductor quantum dots embedded in a topological add-drop filter. The filter is imprinted within a valley-Hall photonic crystal (PhC) membrane and comprises a resonator evanescently coupled to a pair of access waveguides. We show that the longitudinal modes of the resonator enable the filter to perform wavelength-selective routing of light, protected by the underlying topology. Furthermore, we demonstrate that for a quantum dot located at a chiral point in the resonator, selective coupling occurs between well-defined spin states and specific output ports of the topological device. This behaviour is fundamental to the operation of chiral devices such as a quantum optical circulator. Our device therefore represents a topologically-protected building block with potential to play an enabling role in the development of chiral integrated quantum photonic circuits.
\end{abstract}

\maketitle

\section{Introduction}

Photonic crystals (PhCs) are a well-established component in nano-photonic circuitry, their sub-wavelength features supporting low-loss routing of light on-chip within a compact device footprint. Recently, the translation of concepts from the field of topological insulators to the photonic domain has provided new mechanisms for nanoscale control and manipulation of light within PhCs. Notably, photonic analogues of the spin-Hall \cite{Wu2015} and valley-Hall \cite{Ma_2016} effects have been developed; significant advantages of these approaches include robust transport of light around tight bends, intrinsic backscatter immunity, and the potential to form chiral light-matter interfaces in combination with embedded quantum emitters. Devices containing topological photonic interfaces have been demonstrated experimentally using both spin- and valley-Hall PhCs, predominantly in the silicon \cite{Parappuratheaaw4137,Shalaev2019} and GaAs \cite{Barik666,Barik_2020,JalaliMehrabad_APL,JalaliMehrabad_Optica,Yamaguchi_2019,Yoshimi:20} material platforms. In particular, we note that bend robustness has been well established \cite{Barik666,Ma_2019,Yamaguchi_2019,He2019,Shalaev2019,Parappuratheaaw4137,Yoshimi:21} and leveraged to form PhC topological ring resonators \cite{Barik_2020,JalaliMehrabad_APL,JalaliMehrabad_Optica,Gu_2021}.
 
Due to their preservation of time-reversal symmetry, both spin- and valley-Hall photonic analogues support degenerate counter-propagating interface modes, unlike their electronic counterparts. Nevertheless, the modes can be rendered unidirectional by spin selection. At locations known as chiral points, the counter-propagating modes of a topological waveguide have orthogonal circular polarisation. A circularly polarised emitter placed at such a point interacts uniquely with the mode with equivalent handedness, and therefore orthogonal circularly polarised dipoles emit in opposite directions; this is the basis of a chiral light-matter interface \cite{Lodahl2017}. Semiconductor quantum dots (QDs), which have sub-nanosecond radiative lifetimes and have been shown to emit single photons with near-transform limited linewidths \cite{Kuhlmann2015,Pedersen_2020}, are a leading `matter' candidate in this regard; chiral interfaces have been demonstrated using QDs coupled to both spin-Hall \cite{Barik666} and valley-Hall \cite{JalaliMehrabad_Optica,Barik_2020} PhC waveguides. These works built upon substantial prior achievements in topologically trivial chiral systems, for instance those using atoms coupled to microresonators \cite{PhysRevLett.110.213604,Rosenblum2016} and QDs embedded in nanobeam \cite{PhysRevLett.110.037402,Coles2016} or glide plane PhC \cite{Sollner} waveguides. 

The field of on-chip topological quantum photonics is seeing rapid progress; for example, manipulation of quantum states of light has been demonstrated using relatively large topological photonic waveguide devices \cite{Tambascoeaat3187,Wang:19,Blanco-Redondo568,Mittal2021}, while at the nanoscale a passive topological PhC device was recently used as the basis for a simple quantum photonic circuit \cite{PhysRevLett.126.230503}. In contrast, the development of topological photonic devices with integrated quantum emitters is still very much in its infancy. Here, we combine the compactness and chirality afforded by topology to demonstrate chiral emission from a QD embedded in a valley-Hall PhC add-drop filter (ADF) comprising a resonator and pair of access waveguides. We first demonstrate that the device supports wavelength-dependent routing of light. Then, we observe chiral emission from a QD coupled spatially and spectrally to a resonator mode. In the latter measurement, photons emitted after recombination of a specific QD spin state are coupled into two of the four output modes of the device; the chiral nature of the light-matter interface results in the orthogonal spin state coupling to the other two output modes. Such a device has potential applications in on-chip routing of light at the single photon level, for instance as a quantum optical circulator \cite{scheucher_2016_quantum}.

\section{Valley-Hall resonator design}

\begin{figure}[b!]
\includegraphics[width=0.7\linewidth]{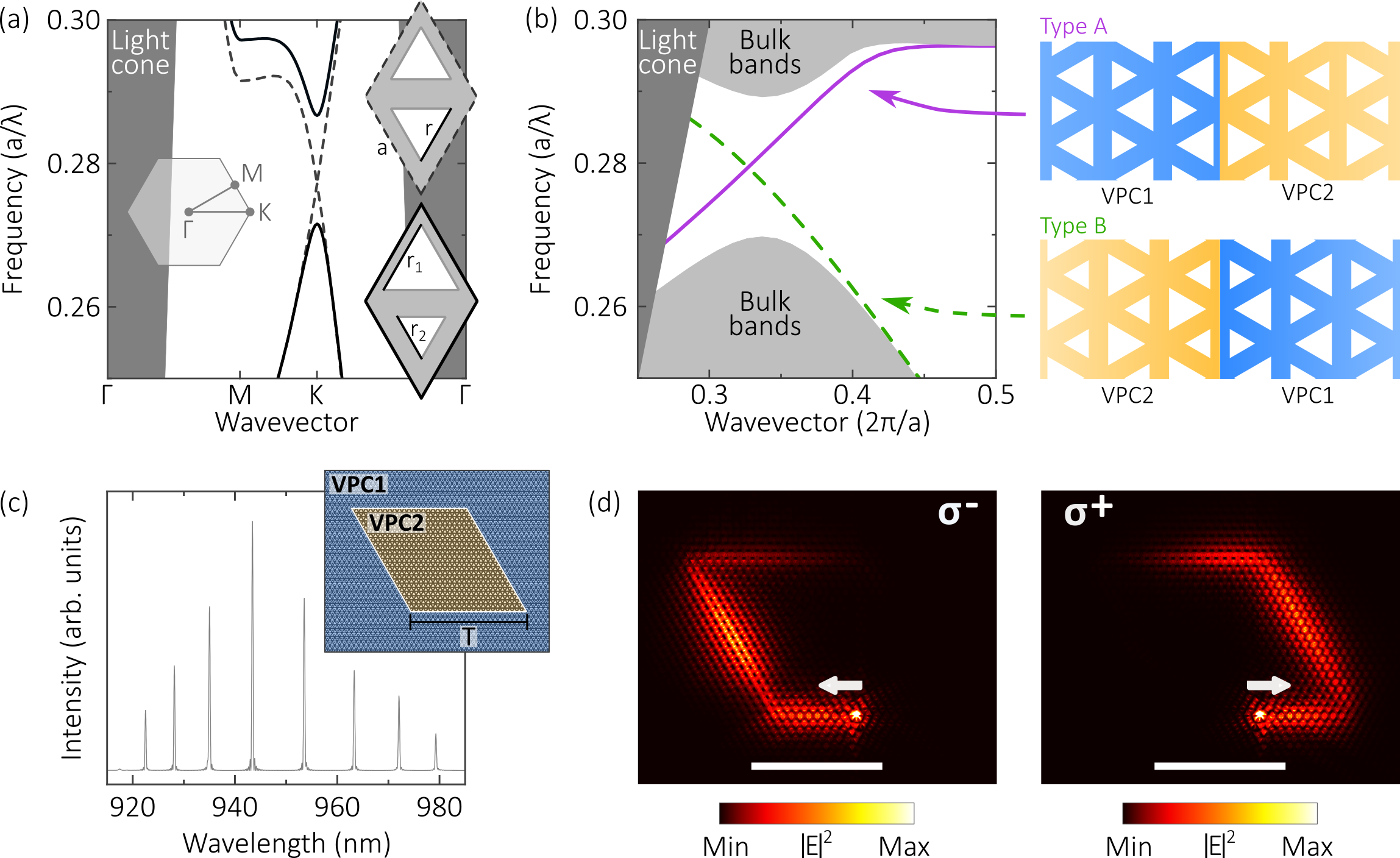}
\caption{(a) Band diagram for a triangular lattice PhC formed using either unperturbed (dashed line) or perturbed (solid line) rhombic unit cells. Insets show (left) the first Brillouin zone and (right) schematics of the unit cells, with equilateral triangles representing air holes in a dielectric membrane. The triangle side lengths are $(r,r_1,r_2)=(0.5a,0.4a,0.6a)$, where $a$ is the lattice constant. The membrane has a thickness of $h=0.64a$ and refractive index $n=3.4$. (b) TE dispersion diagram for two possible zigzag interfaces formed by inverting the unit cells in one half of the PhC. These are labelled type A and type B in the accompanying schematics. (c) Simulated mode spectrum for a rhombic ring resonator formed using the topological zigzag interface. The resonator has a side length of $T=20$ unit cells and lattice period $a=266$nm. A schematic of the resonator is shown in the inset. (d) Simulated, spatially resolved electric field intensity when a resonator mode is excited by a dipole positioned at a chiral point on the interface. The dipole is either (left) $\sigma^-$ or (right) $\sigma^+$ polarised. The electric field intensities are averaged over the first 200fs of simulation time. Arrows indicate the direction of light propagation in each case. Scale bars $4\mu$m.}
\label{fig1}
\end{figure}

Our integrated nano-photonic device is configured within a two-dimensional PhC which has valley-Hall-type topology. The rhombic unit cell of the PhC comprises two equilateral triangular apertures, which are formed within a thin, free-standing dielectric membrane. We evaluate the transverse electric (TE) band structure of the PhC using the freely available MPB software package \cite{Johnson2001:mpb}. When the apertures are of equal size, the TE band structure exhibits a Dirac cone at the $K$ point (see Fig.~\ref{fig1}a). Expanding one aperture while shrinking the other, however, leads to the opening of a topological bandgap at the $K$ point, also shown in Fig.~\ref{fig1}a. We can then take advantage of the topological bulk-edge correspondence to realise within the PhC a waveguide which supports a topologically non-trivial optical mode. To achieve this, we break the inversion symmetry of the PhC by inverting the unit cells in one half of the PhC, creating a zigzag interface at the boundary between the two regions comprising inverted (labelled VPC1) and uninverted (VPC2) unit cells, respectively. Two possible zigzag interfaces can be formed in this way, characterised by the proximity of either the large or small triangular apertures at the boundary of VPC1 and VPC2 unit cells. We refer to these as type A and type B interfaces, respectively, as shown in Fig.~\ref{fig1}b. The associated dispersion diagram shows that both interfaces support a single guided TE mode, which extends across 100\% ($\sim84\%$) of the bandgap for the type A (B) interface.

Two defining characteristics of valley-Hall topologically non-trivial optical modes are (i) their broadband ability to navigate sharp corners without experiencing backscatter (due to the suppression of inter-valley scattering) and (ii) the helical nature of the modes. These properties are critical to the operation of the photonic device reported here. To demonstrate the backscatter protection afforded by topology to the waveguide mode, we use finite-difference time-domain (FDTD) simulations \cite{Lumerical} to probe the optical properties of a rhombus-shaped ring resonator, formed by embedding VPC2 unit cells inside a larger VPC1 unit cell matrix, as shown in Fig.~\ref{fig1}c. In this structure, a bend with 60 (120) degree internal angle connects interfaces of differing (the same) type, giving a maximum resonator spectral bandwidth of $\sim84\%$ of the bandgap, following from the discussion above. The device parameters used in Fig.~\ref{fig1}c are chosen to enable operation in the near infra-red (NIR), compatible with the highest quality QDs and the spectral region in which our later experiments are conducted.

The resulting mode spectrum reveals clear longitudinal modes lying within the topological bandgap ($\sim920-980$nm), with Q factors of the order of $10^5$ for a resonator side length of T=20 unit cells. Significantly, the predicted device performance is broadband in nature, indicating that topological protection overcomes the wavelength dependence commonly observed in light transmission through topologically trivial PhC waveguide bends. Further evidence of topological protection can be found in Fig.~\ref{fig1}d, in which we excite the resonator using a circularly polarised dipole source. By suitable positioning of the dipole at a chiral point (see Supplementary Information section S1) a single unidirectional mode is excited, which travels either clockwise (CW) or counter-clockwise (CCW) around the ring, depending on the handedness of the source polarisation. The mode is seen to navigate the resonator corners smoothly and without backscattering. This simulation also serves to highlight the second key element of our device: the potential to realise a unidirectional light-matter interaction between a resonator mode and an embedded quantum emitter, which ultimately enables spin-dependent routing of light on-chip.

\section{Topological add-drop filter}

\begin{figure}[b!]
\includegraphics[width=0.35\linewidth]{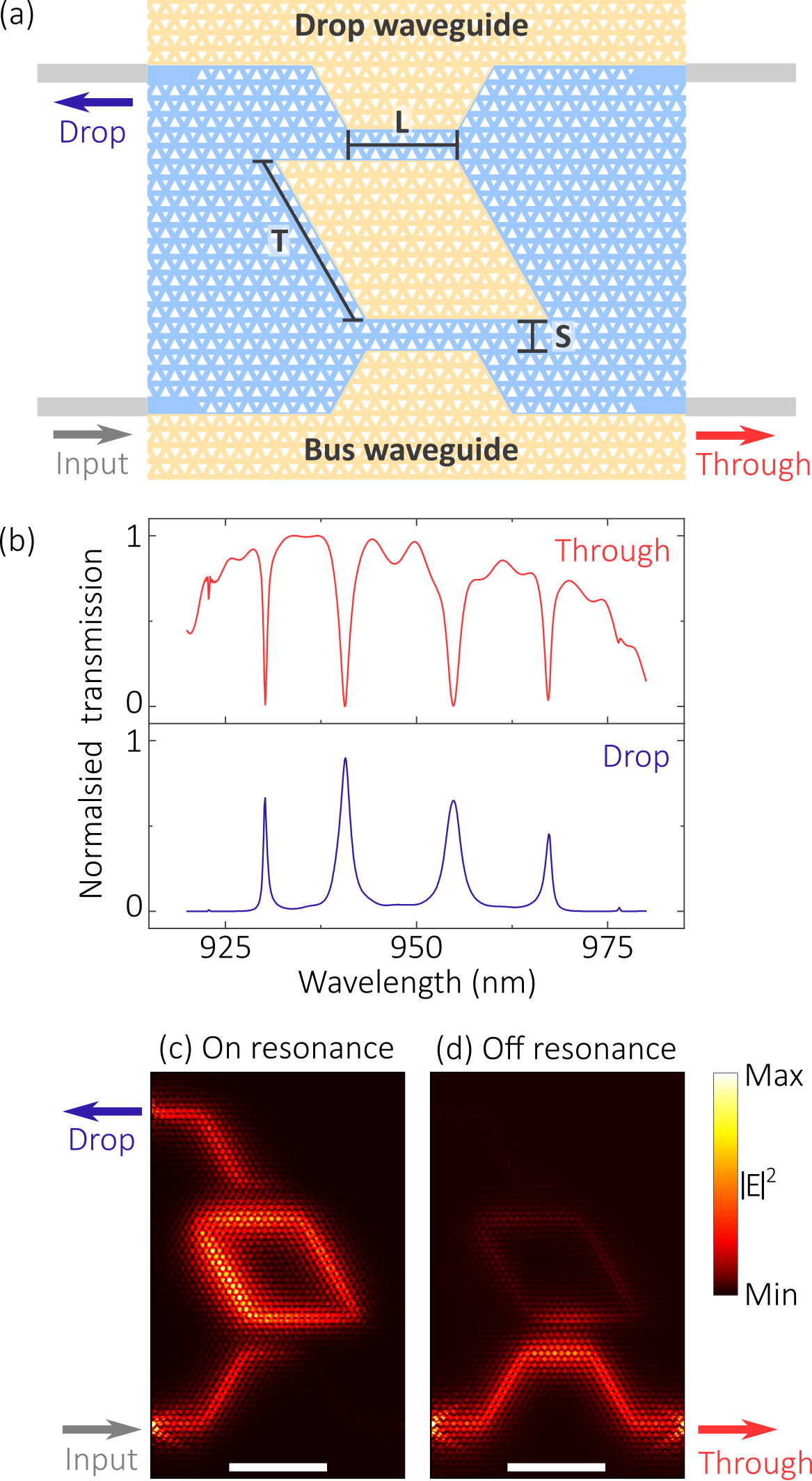}
\caption{(a) Schematic of the topological ADF, with PhC unit cells colour coded as in Fig.~\ref{fig1}b by orientation (inverted (VPC1) or uninverted (VPC2)). (b) Simulated transmission in the through (lower right) and drop (upper left) directions for an ADF with T=16, S=6 and L=12 unit cells, respectively. The data is normalised to the maximum through transmission. Losses in the simulation stem predominantly from the incomplete optimisation of the nanobeam-photonic crystal interfaces. (c) Time-averaged electric field intensity for light input on resonance with the longitudinal mode at 955nm. (d) Time-averaged electric field intensity when the input is off resonance (969nm). Scale bars in (c,d) are $4\mu$m.}
\label{fig2}
\end{figure}

To enable integration of the resonator within scalable nanophotonic circuits, we introduce parallel `bus' and `drop' waveguides in close proximity to opposing sides of the rhombus-shaped resonator, forming an ADF. The filter is shown schematically in Fig.~\ref{fig2}a. Evanescent coupling between waveguides and resonator in this geometry enables light to enter (exit) the resonator from (into) the waveguides. Geometrically, the coupling strength is dependent on the waveguide-resonator coupling length \textit{L} and separation \textit{S}. In our structure, bends in each waveguide allow these parameters to be decoupled, providing increased flexibility in device design. The effect of changing $S$ is investigated in Supplementary Information section S2.

The optical properties of the device are demonstrated using an FDTD simulation in which broadband light is injected into the bus waveguide, and the power subsequently transmitted through the bus waveguide or coupled into the drop waveguide is monitored. The simulation results are summarised in Fig.~\ref{fig2}b-d. In the through direction, the transmission envelope represents the bandwidth in which the type A and B interface modes overlap spectrally. Within this bandwidth clear dips in transmission are observed at wavelengths which correspond to the longitudinal modes of the resonator. The dips occur due to destructive interference between the mode in the bus waveguide and light coupled back into this waveguide from the resonator. Corresponding peaks are seen in the drop channel signal, showing that light is transferred from the input waveguide to the drop waveguide when resonant with the ring modes. The degree to which the input light is dropped on resonance depends on the waveguide-resonator coupling strength, with critical coupling occurring for $S\sim4-6$ unit cells (mode dependent). For the device simulated here, the loaded Q factors lie in the range 400-1600. Off-resonance, the dropped signal is strongly suppressed. Note that the dropped signal propagates in the opposite direction to the input signal due to a combination of valley momentum conservation and the use of both type A and B interfaces in the device. This is examined in detail in Supplementary Information section S3.

\begin{figure}[b!]
\includegraphics[width=0.6\linewidth]{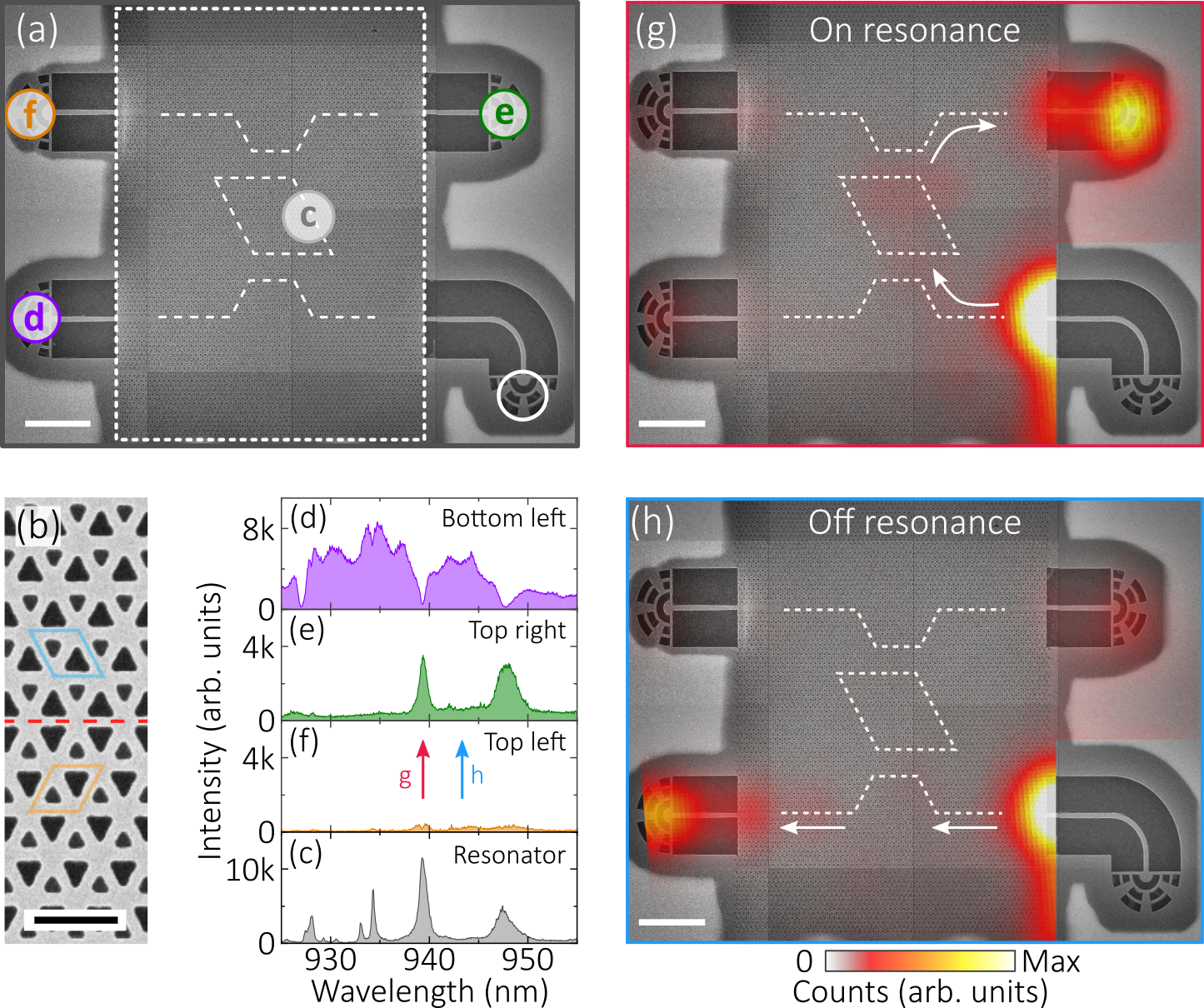}
\caption{(a) SEM image of a topological ADF, with waveguide-resonator separation S=6 unit cells. The topological interfaces forming the waveguides and resonator are highlighted by dashed white lines. (b) Higher magnification SEM image of a type A interface (red dashed line). Unit cells on either side of the interface are outlined. Scale bar 500nm. (c) PL spectrum acquired by exciting the QD ensemble at the resonator interface and collecting emission from the lower right OC (open white circle in (a)). (d-f) PL spectra acquired by exciting the QD ensemble in the lower right OC and collecting emission independently from the other three OCs. The collection position in each case is given in (a). (g,h) Integrated PL intensity as a function of collection position, overlaid on an SEM image of the device. The excitation is fixed above the lower right OC. The integration is taken over (g) 939.3nm to 939.5nm and (h) 943nm to 944nm, respectively. The waveguide and resonator interfaces are identified using white dashed lines, while arrows provide a guide to the the direction of light transmission. Data from above the excitation location (lower right corner) has been omitted, and the zero of the colour scale has been made transparent to aid visualisation of the device.  Scale bars in (a,g,h) are 4$\mu$m.}
\label{fig3}
\end{figure}

Experimentally, we fabricate the topological ADF within a nominally 170nm-thin, \textit{p-i-n} diode GaAs membrane, which contains a single layer of InAs QDs within the diode's intrinsic region. A $1.15\mu$m-thick AlGaAs sacrificial layer beneath the membrane is removed to create a free-standing structure. Details of the fabrication procedure can be found in the Methods. A scanning electron microscope (SEM) image of a representative device is shown in Fig.~\ref{fig3}a. Note that each output of the ADF is coupled to a nanobeam waveguide, which is terminated using a grating outcoupler (OC) for efficient coupling of light into free space optics. A higher magnification SEM image of a type A interface is shown in Fig.~\ref{fig3}b.

To demonstrate the basic operation of the ADF, we inject broadband light into one waveguide via high-power non-resonant excitation of the QD ensemble, and monitor the transmission from the other OCs. We also evaluate the mode spectrum of the resonator itself in a  separate measurement, by generating ensemble photoluminescence (PL) from a fixed position along the resonator interface. The resulting spectra are shown in Fig.~\ref{fig3}(c-f). We first note that the resonator longitudinal modes are broadened due to evanescent coupling into the two waveguides. Quality (Q) factors of 1100 and 440 obtained for the modes at 939nm and 947nm, respectively, in good agreement with the range of values obtained from simulation.

Considering now the signal transmitted through the bus waveguide (Fig.~\ref{fig3}d), we observe several strong dips in transmission which are resonant with the longitudinal modes of the resonator. Indeed, for three of the four modes the transmission is almost completely inhibited. (Note that the spectral envelope in the transmission measurement is governed by the emission from many different QDs, explaining the off-resonant, wavelength-dependent intensity.) In the drop direction (Fig.~\ref{fig3}e), corresponding peaks are observed for the modes at 939nm and 947nm, showing that at these wavelengths, light is coupled from bus to drop waveguide via the resonator. Importantly, we see only very weak emission from the other end of the drop waveguide (Fig.~\ref{fig3}f), demonstrating suppression of backscatter in the topological resonator.

Next, we focus our attention on the mode centered at $\sim939$nm, resonant with which the transmission of the bus waveguide approaches zero (suggestive of critical coupling). We step the collection spot in an x-y grid across the device while keeping the excitation laser fixed above the lower right OC, and acquire a PL spectrum at each collection position. The resulting data is spectrally filtered such that it corresponds to either on or off resonance with the longitudinal mode at $\sim939$nm. Spatial maps giving the integrated PL signal are shown in Fig.~\ref{fig3}(g,h). On resonance, light is detected predominantly from the drop (top right) OC, showing that it is coupled through the resonator. Conversely, when off resonant the largest signal is from the left hand OC of the bus waveguide, bypassing the resonator. Scattering is also observed at the nanobeam-topological waveguide interfaces, which were not optimised in this device. More significantly, minimal scattering is observed from above the bends in the topological interface, testifying to their quality. A complementary measurement in which the excitation location was scanned whilst PL emission was collected from a single fixed OC is presented in Supplementary Information section S4.

\begin{figure}[t!]
\includegraphics[width=0.8\linewidth]{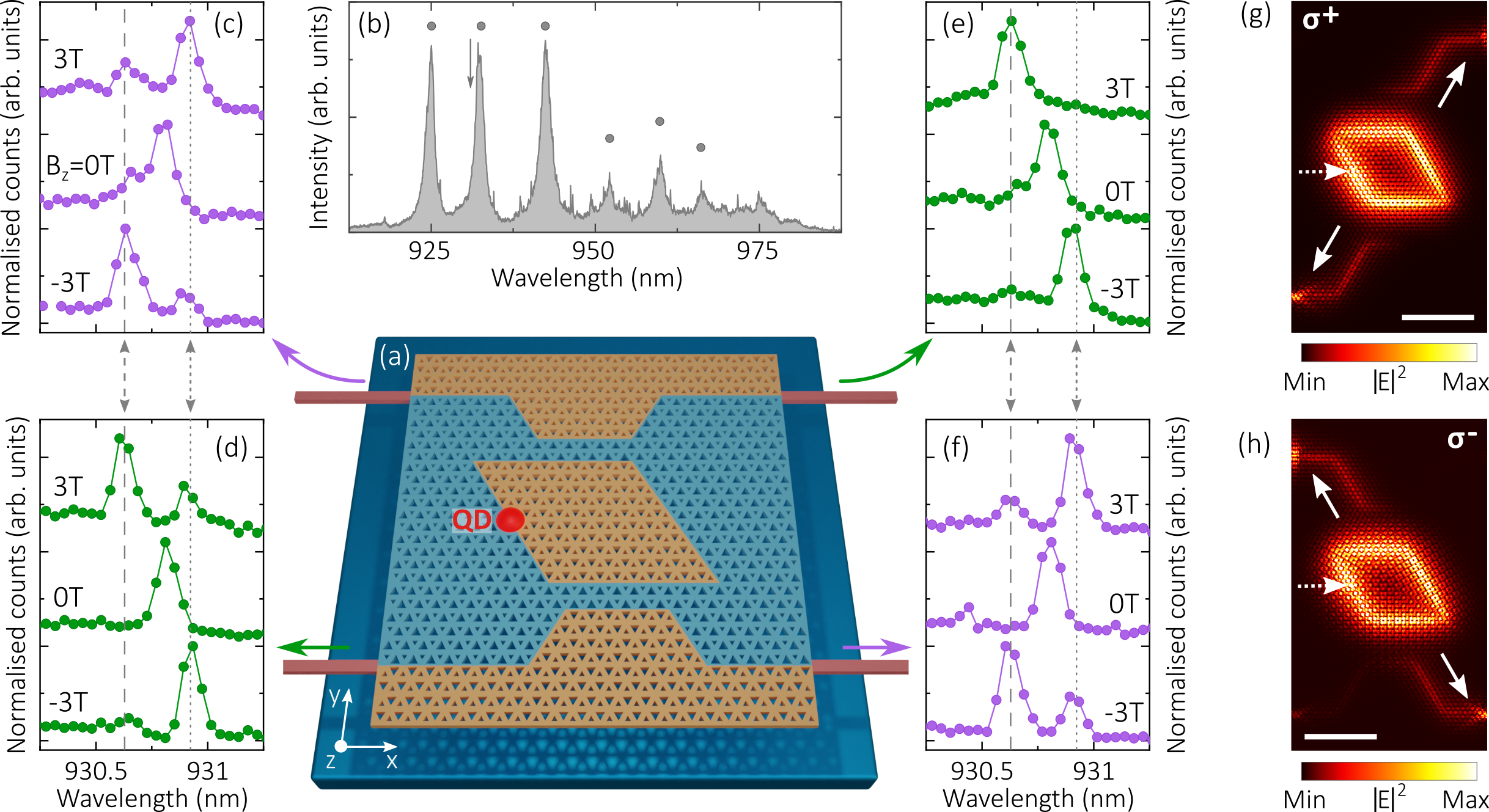}
\caption{(a) Schematic of the ADF, with a QD embedded at the resonator interface. Each of the four nanobeam waveguides is terminated with a grating outcoupler (OC), not shown. (b) PL spectrum acquired from an OC under high-power excitation of the resonator interface. The longitudinal resonator modes are marked by filled circles. The device parameters are (T,S,L) = (16,10,11) unit cells, respectively. (c-f) Low power PL spectra for a single QD located at the resonator interface and with an optical transition near-resonant with a longitudinal mode (see arrow in (b)). The spectra are acquired from the OCs in the (c) top left, (d) lower left, (e) top right and (f) lower right of the device, respectively. Data is shown for three different magnetic field strengths, $B_z=-3$T, 0T and 3T, respectively. Dashed and dotted lines indicate the approximate wavelengths of the two QD transitions under non-zero magnetic field. (g-h) Simulated, time-averaged electric field intensity in the plane of the device, for a (g) $\sigma^+$ or (h) $\sigma^-$ polarised dipole placed at a chiral point at the resonator interface (approximate location given by the dashed arrow). Solid arrows indicate the dominant coupling directions from of the resonator in each case. Scale bars 4$\mu$m.}
\label{fig4}
\end{figure}

Having demonstrated the basic function of the device, we now consider a second operational paradigm, in which we utilise single QDs embedded at the resonator interface. We first perform a Hanbury-Brown and Twiss (HBT) measurement on photons emitted from a QD which is coupled spectrally and spatially to a resonator mode. In so doing, we collect light into separate optical fibers from either end of one bus waveguide, therefore performing the HBT beam-splitting operation on-chip (see Supplementary Information section S5). We obtain a $g^{(2)}(0)$ of 0.14 after correcting for background emission from other QDs coupled to the same resonator mode, demonstrating the single photon nature of the QD emission.

Next, we investigate the chiral behaviour of a single QD in a similar device. A schematic of the structure and the measured resonator spectrum are shown in Fig.~\ref{fig4}a and Fig.~\ref{fig4}b, respectively. The QD is excited non-resonantly and PL emission is subsequently detected from all four ports of the ADF. The resulting spectra as a function of magnetic field are shown in Fig.~\ref{fig4}c-f. Clear routing of light dependent on the spin state of the QD transition is observed. First, we consider the spectra obtained at $B_z=0$T, for which the two spin states of the QD transition are degenerate, and note that a single PL emission line is observed from all output ports. Due to the statistical emission from both spin states of the QD, the source in this case is effectively unpolarised and therefore excites both CW and CCW resonator modes, which subsequently decay equally into the four output ports. 

However, upon application of a magnetic field in the Faraday geometry, the Zeeman effect leads to spin splitting, giving rise to two non-degenerate, orthogonal, circularly polarised QD transitions. For a suitably positioned emitter, spin-momentum locking at the topological interface ensures that the direction of emission from the QD is dependent on the spin state. One of the spin states therefore couples to the CW resonator mode, while the other spin state couples to the CCW mode. The modes subsequently decay directionally into the two waveguides. Valley-momentum conservation results in a single resonator mode coupling to diagonally opposing ports of the ADF, and therefore a single QD spin state also couples in the same manner in our device. For instance, at a magnetic field of 3T the dominant emission line measured from the top left and lower right OCs is at longer wavelength, while the opposite is true for the other two OCs. When the sign of the magnetic field is flipped, the situation is reversed, with the shorter wavelength peak becoming dominant in the top left and lower right data. Thus, the device acts as a bidirectional chiral router for each spin state of the QD. FDTD simulations for $\sigma^{+}$ and $\sigma^{-}$ dipoles placed at the resonator interface support the experimental observation (see Fig.~\ref{fig4}(g,h)). Additional experimental data is shown in Supplementary Information section S6 for a second QD coupled to a different mode of the same device, demonstrating the broadband potential of the structure.

\section{Discussion}

We have demonstrated an integrated topological add-drop filter operating in the optical domain. The filter consists of a compact resonator coupled to a pair of access waveguides, defined within a valley-Hall topological PhC. Characterisation of the device using broadband emission from the QD ensemble indicates that near-critical coupling between the waveguides and resonator is achieved, allied with suppressed scatter, both in and out of plane. Chiral emission from a QD embedded within the resonator is also demonstrated. One spin state of a QD transition is shown to couple into two of the four output ports of the device, with the orthogonal spin state coupling to the other two output ports. 

In future, optimisation of the device to increase the intrinsic (unloaded) resonator Q factor could enable broadband Purcell enhancement \cite{PhysRevApplied.16.014036} of the chiral light-matter interaction. This might be achieved, for instance, by passivation of the semiconductor surface to suppress surface-related losses \cite{Guha:17,Kuruma_2020,PhysRevApplied.16.014036}. Scale up could then be envisioned; for example, cascaded chiral resonators could be used for the transport of entangled states on-chip \cite{PhysRevResearch.2.013369}. Addressing the QD resonantly in the coherent scattering regime \cite{Rattenbacher_2019,brooks2021integrated} could ultimately allow for the realisation of a topologically protected, integrated quantum optical circulator \cite{scheucher_2016_quantum}.

\section{Methods}

\subsection{Device fabrication}
The device layers were grown on a semi-insulating (100) GaAs substrate using molecular beam epitaxy. In order of deposition, the layers are: 1$\mu$m Al$_{0.6}$Ga$_{0.4}$As, 30nm n-GaAs, 50nm Al$_{0.3}$Ga$_{0.7}$As, 5nm i-GaAs, InAs QDs, 5nm i-GaAs, 30nm Al$_{0.3}$Ga$_{0.7}$As and 50nm p-GaAs.

Nanophotonic devices were fabricated using standard lithography and wet/dry etching techniques. A 120nm-thick SiO\textsubscript{x} hardmask was deposited on the wafer using plasma enhanced chemical vapour deposition. This was followed by spinning of an electron-beam-sensitive resist (CSAR). The devices were subsequently patterned using 50kV electron beam lithography (Raith Voyager) and then etched into the hardmask and epitaxial layers using reactive ion etching (RIE) and inductively coupled plasma RIE, respectively. The hardmask and AlGaAs sacrificial layer were removed using a hydrofluoric acid wet etch.

\subsection{Experimental methods}

The sample was mounted in a superconducting magnet cryostat (Cryo Industries of America) operating at 4.2K. PL measurements were undertaken using non-resonant laser excitation at 770nm (M Squared SolsTiS). For the creation of spatial PL maps, the excitation or collection spot was rastered across the sample using a motorized mirror in the microscope excitation or collection path, respectively. The use of a relay lens pair ensured that the laser and collection spots remained well-focused across the full scanning range during this process.

For autocorrelation (HBT) measurements, the sample was excited using an 808nm diode laser (Thorlabs CPS808). The signal was collected independently from either end of one bus waveguide, and coupled into separate optical fibers. Two 0.75m monochromators (Princeton Instruments SP-2750) were used to filter the output (filter bandwidth $\sim 0.1$nm) which was then detected using two avalanche photodiodes (Excelitas SPCM) with a convolved instrument response time of $\sim 700$ps. Photon arrival times were recorded and time-correlated using two channels of a time tagger (Swabian Instruments TimeTagger Ultra).

\section*{Author Contributions}

M.J.M. designed the photonic structures, which R.D. fabricated. E.C. and P.K.P. grew the sample. M.J.M., A.P.F. and N.J.M. carried out the measurements and simulations. L.R.W. and M.S.S. provided supervision and expertise. A.P.F. wrote the manuscript, with input from all authors. 

\section*{Acknowledgements}

This work was funded by the Engineering and Physical Sciences Research Council (EPSRC) (Grant No. EP/N031776/1).

\bibliography{main.bib}
\end{document}


\title{Supporting Information for: A chiral topological add-drop filter for integrated quantum photonic circuits}

\author{M. Jalali Mehrabad}
\email[]{mjalalimehrabad1@sheffield.ac.uk}
\affiliation{Department of Physics and Astronomy, University of Sheffield, Sheffield S3 7RH, UK}

\author{A.P. Foster}
\email[]{andrew.foster@sheffield.ac.uk}
\affiliation{Department of Physics and Astronomy, University of Sheffield, Sheffield S3 7RH, UK}

\author{N.J. Martin}
\affiliation{Department of Physics and Astronomy, University of Sheffield, Sheffield S3 7RH, UK}

\author{R. Dost}
\affiliation{Department of Physics and Astronomy, University of Sheffield, Sheffield S3 7RH, UK}

\author{E. Clarke}
\affiliation{EPSRC National Epitaxy Facility, University of Sheffield, Sheffield S1 4DE, UK}

\author{P.K. Patil}
\affiliation{EPSRC National Epitaxy Facility, University of Sheffield, Sheffield S1 4DE, UK}

\author{M.S. Skolnick}
\affiliation{Department of Physics and Astronomy, University of Sheffield, Sheffield S3 7RH, UK}

\author{L.R. Wilson}
\affiliation{Department of Physics and Astronomy, University of Sheffield, Sheffield S3 7RH, UK}

\begin{abstract}
This file contains supporting information for the paper ``A chiral topological add-drop filter for integrated quantum photonic circuits". 
\end{abstract}

\maketitle
\tableofcontents

\newpage
\section{Mode profiles and chiral spatial maps for the topological interface.}

\begin{figure}[h!]
\includegraphics[width=0.8\linewidth]{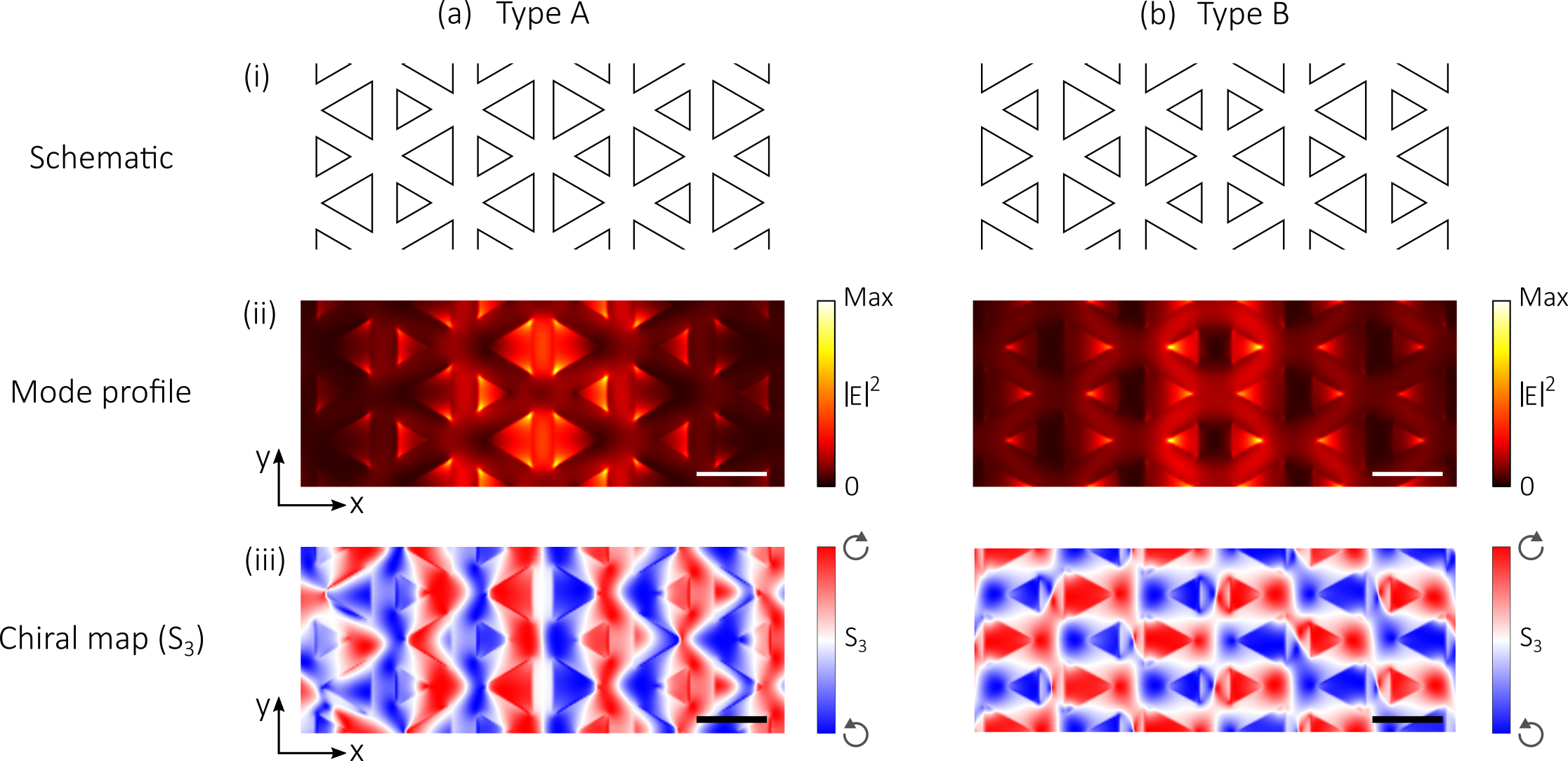}
\caption{Mode profiles and chiral maps for (a) type A and (b) type B valley-Hall zigzag topological interfaces, respectively. In each case (i) shows a schematic of the interface, (ii) gives the electric field intensity spatial map of the guided mode and (iii) shows the (normalised) $S_3$ Stokes parameter, giving the degree of circular polarisation of the mode as a function of $(x,y)$ position. Scale bars in (b,c) are 200nm.}
\label{figSI1}
\end{figure}

Fig.~\ref{figSI1} shows the electric field intensity distribution of the guided mode for the type A and type B valley-Hall zigzag topological interfaces, as introduced in Fig. 1 of the main text. Also shown in each case is the normalised Stokes $S_3$ parameter (degree of circularity) for the same cross-section of the waveguide. This is otherwise known as a chiral map, with regions of $|S_3|\to 1$ suppporting chiral coupling of an embedded emitter with circularly polarised transition dipole moment. Using the fields obtained from finite-difference time-domain (FDTD) simulations, the $S_3$ parameter was obtained from the expression

\begin{equation}
    S_3=\frac{-2\textrm{Im}(E_xE_y^{*})}{\lvert E_x\rvert^2+\lvert E_y\rvert^2},
\end{equation}
where $S_3$, $E_x$ and $E_y$ are dependent on $x$ and $y$, with $z=0$ (corresponding to the centre of the PhC membrane).

\newpage
\section{Control of ADF coupling strength using waveguide-resonator separation.}

\begin{figure}[h!]
\includegraphics[width=0.9\linewidth]{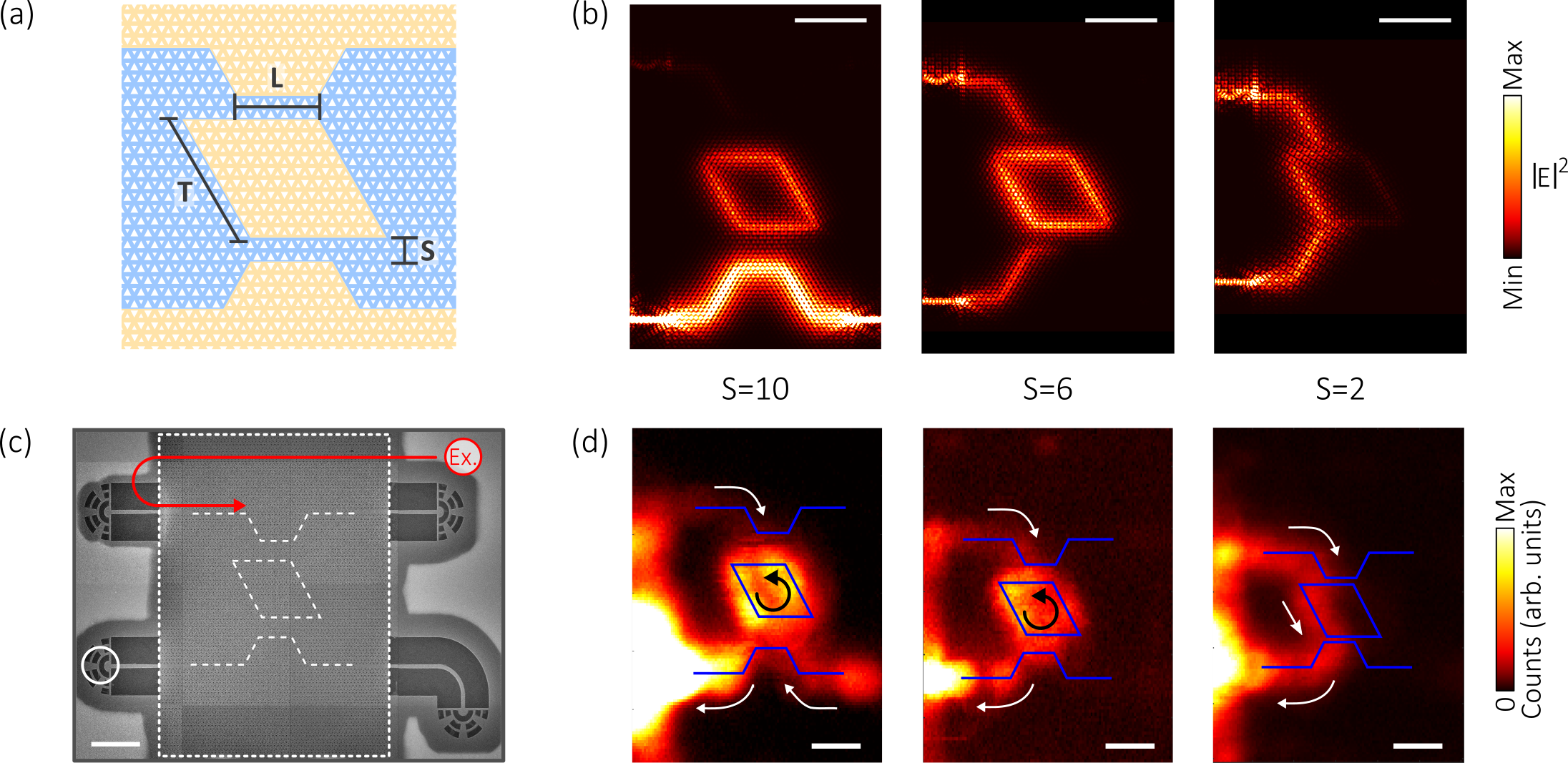}
\caption{Control of ADF coupling strength by changing the waveguide-resonator separation. (a) Schematic of the ADF. (b) Time-averaged electric field intensity spatial maps for three ADFs with different waveguide-resonator separation (from left to right, S=10, 6 and 2 unit cells, respectively). In each case, light which is resonant with a single longitudinal mode of the resonator is injected into the bottom left waveguide.  (c) SEM of an ADF. Ex. - excitation spot location. (d) Integrated PL intensity spatial maps for devices with (left to right) S=10, 6 and 2 unit cells. PL is collected from the OC terminating the lower left output port, while the excitation laser spot is rastered across the device (indicated by the red arrow in (c)). In each case, the integration is taken over a single longitudinal mode at $\sim1000$nm. Solid blue lines give the positions of the waveguides and resonator. White and black arrows indicate the dominant direction of power flow. Scale bars in (b,c) are 4$\mu$m.}
\label{figSI3}
\end{figure}

Here, we demonstrate control of the the add-drop filter (ADF) coupling strength at the design stage. To do so, we vary the waveguide-resonator separation $S$ while keeping the parameters $T$ and $L$ constant (see Fig.~\ref{figSI3}a). By setting $S=10, 6 \mathrm{\ or\ } 2$ unit cells, the ring resonator and the bus and drop waveguides can be weakly, critically or over coupled, respectively. This is demonstrated in Fig.~\ref{figSI3}b, in which FDTD simulations are used to evaluate the time-averaged electric field intensity when light, resonant with the same longitudinal mode in each case, is injected into the drop waveguide of the ADF.

Next, we experimentally investigate structures with these same parameters using integrated photoluminescence (PL) intensity raster scans; the results are shown in Fig.~\ref{figSI3}d. For each of the three devices, we position the collection spot above the lower left outcoupler (OC) and raster scan the excitation laser across the device, as shown in Fig.~\ref{figSI3}c (this is the reverse of the collection raster scan used in the main text). By virtue of the spatially distributed nature of the QD ensemble, this approach effectively provides a user-positioned broadband light source with which to probe the optical response of the device. We integrate the measured PL intensity over a bandwidth corresponding to the same resonator mode in each case. 

For $S=10$, light from the drop waveguide weakly couples to the ring, and we see emission which originates from both waveguides, and from within the resonator. For the $S=6$ case, in which the the mode is near-critically coupled, light is seen to couple through the resonator from the drop (upper) waveguide, while light from the other ends of the drop or through waveguides is not detected. This light is instead coupled via the resonator to the OC on the right hand side of the bus waveguide, in a mirror image of the data shown here. Finally, when $S=2$, the resonator is overly coupled to the waveguides. In this scenario, the light coupled from the drop waveguide to the ring subsequently couples quickly to the bus waveguide.

Note that the approaches used to obtain the simulation data in Fig.~\ref{figSI3}b and experimental data in Fig.~\ref{figSI3}d are not directly equivalent; nevertheless, in combination they clearly demonstrate how changing the parameter $S$ changes the resonator-waveguide coupling strength.

\newpage
\section{Valley momentum conservation in the ADF.}

\begin{figure}[b!]
\includegraphics[width=\linewidth]{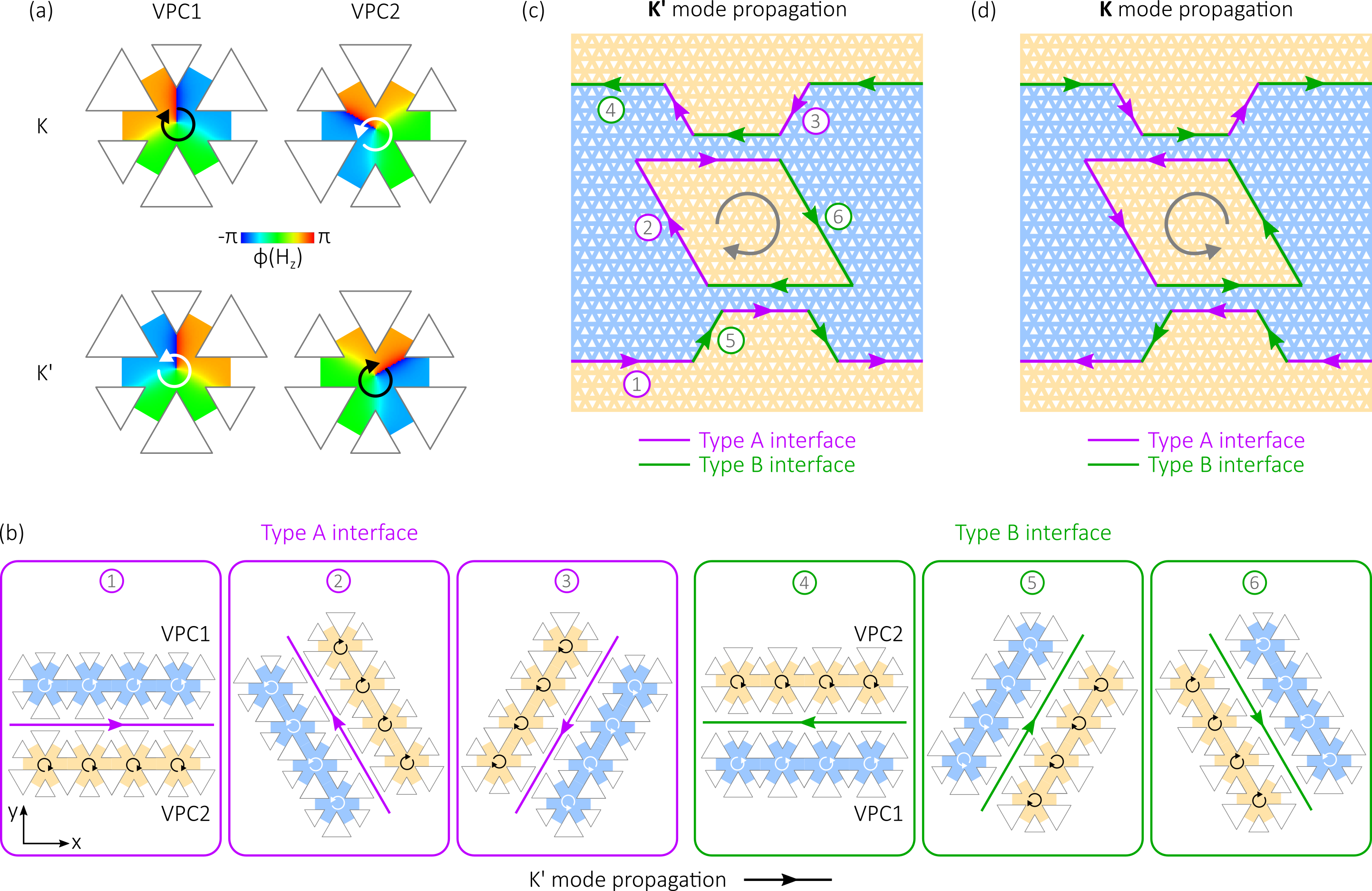}
\caption{(a) Phase vortices for the $H_z$ field at the $K$ and $K'$ points of the lowest frequency TE band, for PhCs formed using either inverted (VPC1) or uninverted (VPC2) unit cells. (b) Schematic showing the direction of light propagation within the $K'$ mode for the six possible combinations of interface type and waveguide orientation in the topological PhC. The phase vortices for the VPC1 and VPC2 unit cells are given by white and black circular arrows, respectively. Straight arrows for the type A and B interface propagation directions are coloured purple and green, respectively.  Waveguides 1, 2 and 3 are equivalent, differing only by rotation; the same is true for waveguides 4, 5 and 6. All arrows are inverted for propagation in the $K$ direction. (c,d) Schematic showing the direction of light propagation within the (c) $K'$ or (d) $K$ mode of an ADF.}
\label{figSI6}
\end{figure}

The direction in which light propagates in the ADF depends on the interface type (A or B) and is dictated by valley momentum conservation \cite{PhysRevLett.126.230503}. This can be understood by considering the phase winding of the $H_z$ field at the $K$ and $K'$ points in the Brillouin zone, and the dependence of the winding direction on the orientation of the PhC unit cell (and therefore the interface type). In Fig.~\ref{figSI6}a we plot the phase of the $H_z$ field for the lowest frequency TE mode at the $K$ and $K'$ points, for both VPC1 and VPC2 PhCs. (The phase was obtained using the freely available MIT Photonic Bands (MPB) software package \cite{Johnson2001:mpb}.) For a PhC formed using VPC1 unit cells, the phase at the $K$ point winds by 2$\pi$ in the clockwise (CW) direction about the centre point. When the unit cells are inverted (i.e. the PhC becomes VPC2-type) the phase rotates in the counter-clockwise (CCW) direction. Conversely, at the $K'$ point the opposite behaviour is observed. 

A type A interface is now formed by positioning a VPC1 PhC above a VPC2 PhC (i.e. separated in the y direction), with the resulting waveguide oriented along the x direction (see schematic 1 in Fig.~\ref{figSI6}b). At the $K'$ point, the field vortices indicate that the interface mode will propagate in the positive x direction. Conversely, for a type B interface the VPC2 PhC is positioned above the VPC1 PhC, and the field vortices are therefore reversed. The $K'$ interface mode in this case propagates in the negative x direction (see schematic 4 in Fig.~\ref{figSI6}b). 

Within the ADF, there are six different combinations of interface type (A or B) and direction (inclined at 0, 60 or 120 degrees to the x axis). However, these can all be traced back to one of the two scenarios outlined above, by a suitable rotation of the waveguide. For instance, the interfaces in schematics 1, 2 and 3 in Fig.~\ref{figSI6}b are equivalent other than by a rotation. The same is true for the interfaces labelled 4, 5 and 6. Using this basis, we indicate on a schematic of the ADF in Fig.~\ref{figSI6}c the direction of propagation for the $K'$ mode. If light is injected into the right travelling mode in the bus (lower) waveguide, a CW resonator mode is subsequently excited; this decays into the left travelling mode in the drop (upper) waveguide. The reversed situation for the $K$ mode is shown in Fig.~\ref{figSI6}d.

Using the above considerations, if light is injected into the $K$ mode in the bus waveguide (i.e. propagating from right to left), transmission into the left travelling $K'$ mode in the drop waveguide should be inhibited, as this necessitates flipping of the $k$-vector. In Fig. 3f of the main text, this suppression of coupling into the `wrong' mode of the experimental drop waveguide is clearly seen.

Note also that topological protection should result in suppression of reflection by the ADF, as this also necessitates flipping of the $k$-vector. An observation along these lines has been made recently in an elastic wave spin-Hall system \cite{10.1093/nsr/nwaa262}, in which topological protection prohibits quasi-spin flips.

\newpage
\section{Excitation PL spatial mapping.}

\begin{figure}[h!]
\includegraphics[width=0.8\linewidth]{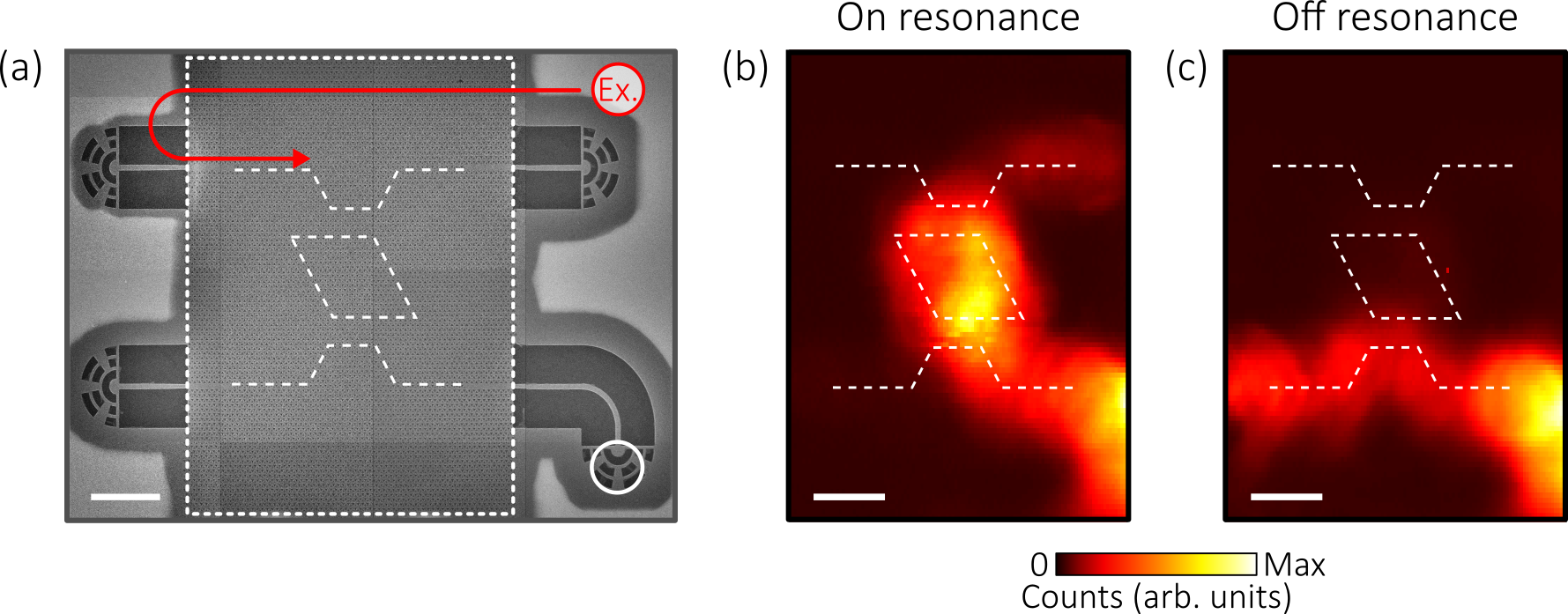}
\caption{(a) SEM of an ADF. Ex. - excitation spot location, which is rastered across the device. (b, c) Integrated PL intensity as a function of excitation position, for the device shown in Figure 3 of the main text. PL is collected from the OC terminating the lower right OC (white circle in (a)), while the excitation laser spot is rastered across the device. The integration is taken over (b) 939.3nm to 939.5nm or (c) 943nm to 944nm. Dashed white lines give the positions of the waveguides and resonator. The area shown in the maps corresponds to the dotted white box in (a). Scale bars in (a-c) are 4$\mu$m.}
\label{figSI4}
\end{figure}

In addition to the collection raster scans shown in in Fig. 3 of the main text, we have performed excitation raster scans on the same device. In this complementary measurement, we position the collection above the bottom right OC and raster scan the excitation laser across the device (the reverse of the collection raster scan shown in the main text). When the PL emission is spectrally filtered such that the emission energy is off-resonance with the mode, the (integrated) PL is seen to originate only from the through waveguide (Fig.~\ref{figSI4}c). In this scenario, the resonator and waveguide modes are decoupled. On resonance, however, light is detected which originates either from one end of the drop waveguide (i.e. coupled through the resonator), or from within the resonator itself (Fig.~\ref{figSI4}b). As the mode is near-critically coupled, light from the other ends of the drop or through waveguides is not detected. This light is instead coupled via the resonator to the OCs on the left hand side of the device, in a mirror image of the data shown here.

\newpage
\section{ADF as a single photon beamsplitter for autocorrelation measurements.}

\begin{figure}[h!]
\includegraphics[width=0.9\linewidth]{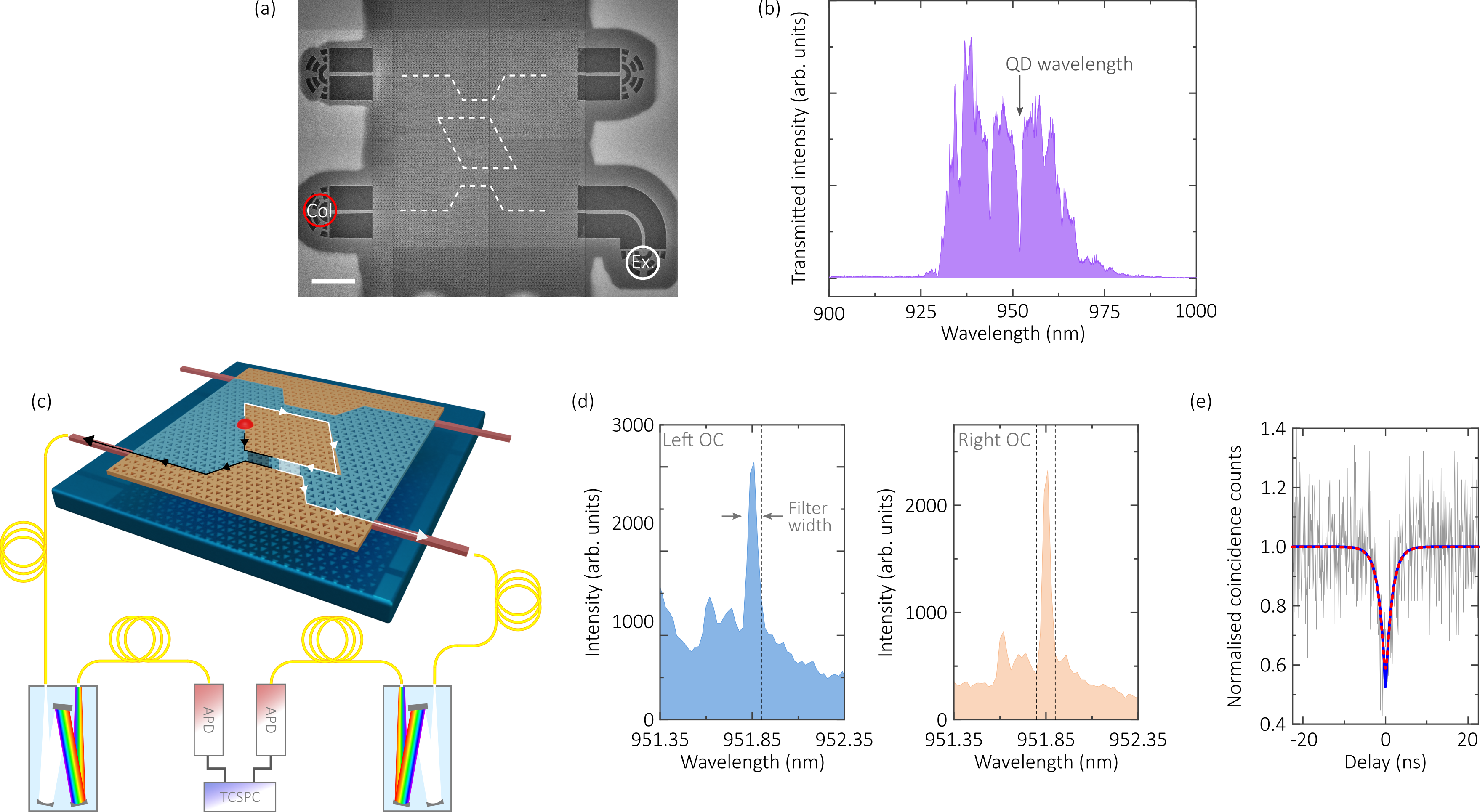}
\caption{(a) SEM of an ADF. Scale bar $4\mu$m. (b) Transmitted PL spectrum acquired from the bottom left OC (red circle in (a)) after excitation of the QD ensemble in the bottom right OC. (c) Schematic of the HBT experimental setup. APD - avalanche photo diode; TCSPC - time correlated single photon counter. (d) PL spectra obtained from the left and right OCs of the waveguide in (c), respectively, when exciting a QD positioned at the resonator interface. The spectra are shown prior to spectral filtering. The QD transition is resonant with a cavity mode (arrow in (b)). (e) Second order auto-correlation measurement for the QD transition in (d). The red dashed line shows a convolved fit to the data, while the blue line shows the deconvolved fit, with $g^{(2)}(0)=0.53$. The bin width is 100ps.}
\label{figSI5}
\end{figure}

Second order autocorrelation measurements (or Hanbury-Brown and Twiss (HBT) measurements) are a standard technique to demonstrate single photon emission from a quantum emitter such as a QD. Typically, one collects photons emitted by the QD and directs them to a pair of single photon detectors via a 50:50 beamsplitter. For a perfect single photon source, correlations between arrival times of the photons show complete suppression at zero time delay. Here, we use the add-drop filter itself as an on-chip beamsplitter. An SEM of the device is shown in Fig~\ref{figSI5}a. First, we identify the spectral position of resonator modes which are coupled to the bus waveguides. We probe the transmission of the lower bus waveguide by exciting ensemble PL from within the bottom right hand OC, detecting the signal from the bottom left hand OC. The resulting transmission spectrum is shown in Fig~\ref{figSI5}b, with several clear dips observed due to waveguide-resonator coupling. 

Next, we identify a QD located within the resonator and coupled to the mode at $\sim 952$nm. Using low power non-resonant excitation ($\lambda_{\mathrm{laser}}=808$nm), we collect PL simultaneously from both ends of the lower bus waveguide, as indicated in the schematic in Fig~\ref{figSI5}c. The QD couples to both CW and CCW resonator modes with similar efficiency, and the modes subsequently decay in either direction along the bus waveguide, thereby realising an approximately 50:50 beamsplitting operation. We filter the signal collected by both fibers using separate monochromators with $\sim 100$pm filter bandwidth, with the signals prior to filtering shown in Fig~\ref{figSI5}d. Note that due to the relatively high density of QDs in the sample, the filtered signal still contains a contribution from QDs other than the target QD transition. We then correlate the arrival times of photons from either collection fiber using a pair of APDs and time tagging electronics. The result of the HBT measurement is shown in Fig~\ref{figSI5}e, with $g^{(2)}(0)=0.53$ after deconvolution of the instrument response. If we then account for the signal to background ratio obtained from the filtered PL spectra, we obtain a corrected value of $g^{(2)}(0)=0.14$, clearly demonstrating the single photon nature of the emission.

\newpage
\section{Broadband nature of chiral coupling in the ADF.}

\begin{figure}[h!]
\includegraphics[width=0.6\linewidth]{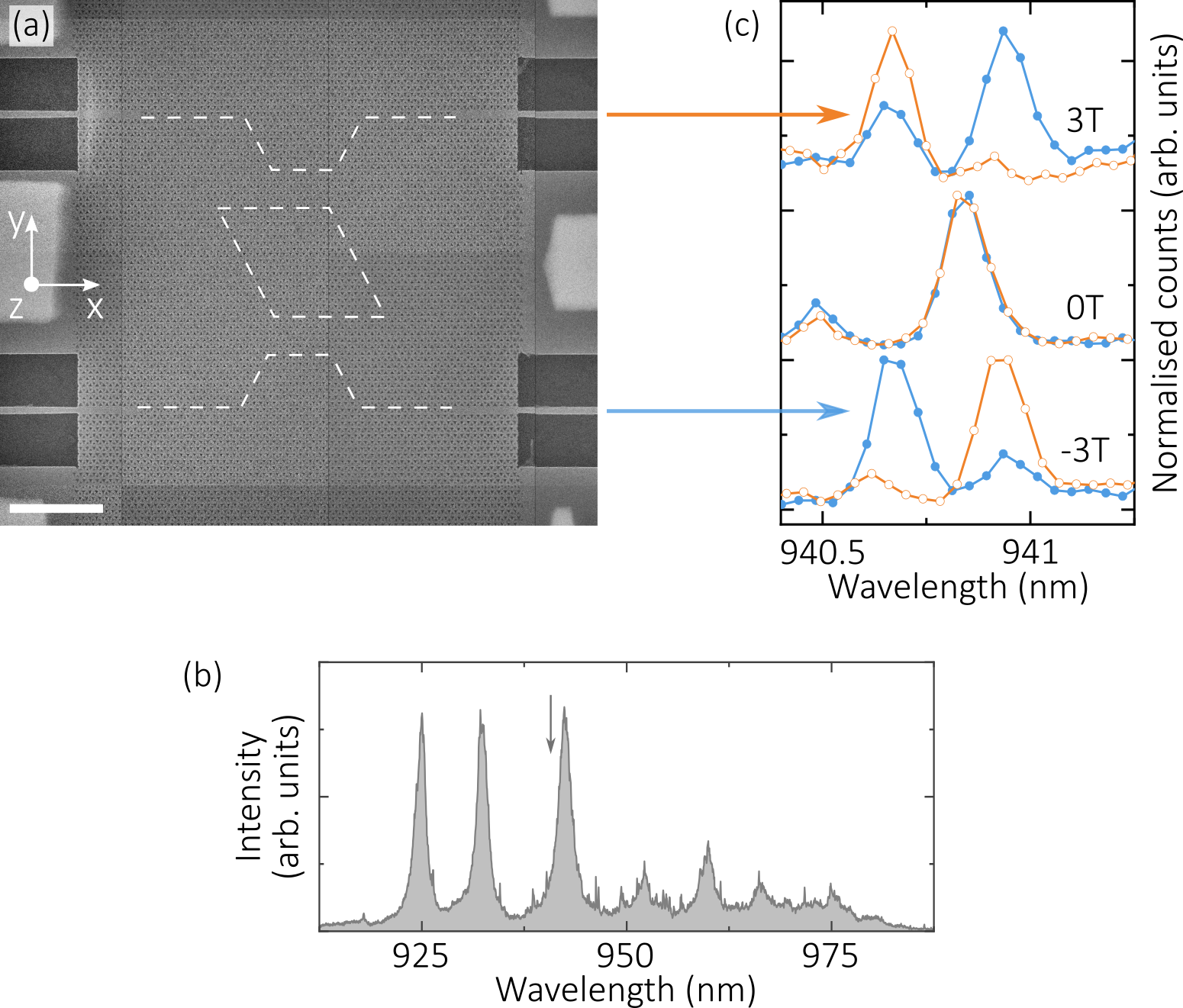}
\caption{SEM image of a representative ADF, with waveguide-resonator separation S=6 unit cells. Note that the measured device has S=10 unit cells. Dotted lines outline the waveguide and the resonator interfaces. Each nanobeam waveguide is terminated with an OC (not shown). Scale bar 4$\mu$m. (b) PL spectrum acquired from an OC under high-power excitation of the resonator interface. (c) Low power PL spectra for a single QD located at the resonator interface and with an optical transition near-resonant with a longitudinal mode (see arrow in (b)). The spectra are acquired from the OCs at the top right (orange line) and bottom right (blue line) of the device, respectively. Data is shown for three different magnetic field strengths, $B_z=-3$T, 0T and 3T, respectively.}
\label{figSI2}
\end{figure}

Fig.~\ref{figSI2} shows chiral coupling of a second QD to the device considered in Fig. 4 of the main text. Most importantly, note that the QD is coupled to a different mode of the resonator in this case. The QD is excited non-resonantly and PL emission is subsequently detected from the two ports on the right hand side of the ADF. The resulting spectra as a function of magnetic field are shown, clearly demonstrating routing of light dependent on the spin state of the QD transition.

\bibliography{main.bib}